\documentclass[osajnl,twocolumn,showpacs,superscriptaddress,10pt]{revtex4-1} 
\usepackage{amsmath,amssymb,graphicx}
\newcommand{\eryso}[0]{Er$^{3+}$:Y$_2$SiO$_5$ }
\newcommand{\Pras}[0]{Pr$^{3+}$:Y$_2$SiO$_5$ }
\newcommand{\yso}[0]{Y$_2$SiO$_5$ }

\newcommand{\dun}[0]{ {\bf D$_1$} }
\newcommand{\ddeux}[0]{ {\bf D$_2$} }
\newcommand{\baxis}[0]{ {\bf b} }

\begin{document}

\title{Quantum memory for light: large efficiency at telecom wavelength}

\author{Juli\'an Dajczgewand}
\author{Jean-Louis Le~Gou\"et}
\author{Anne Louchet-Chauvet}
\author{Thierry Chaneli\`ere}\email{Corresponding author: thierry.chaneliere@u-psud.fr}
\affiliation{Laboratoire Aim\'e Cotton, CNRS-UPR 3321, Univ. Paris-Sud, B\^at. 505, 91405 Orsay cedex, France}

\begin{abstract}We implement the ROSE protocol in an erbium doped solid, compatible with the telecom range. The ROSE scheme is an adaptation of the standard 2-pulse photon echo to make it suitable for a quantum memory. We observe an efficiency of 40\% in a forward direction by using specific orientations of the light polarizations, magnetic field and crystal axes.
\end{abstract}


\maketitle 


The use of erbium doped materials has revolutionized fiber-optic communications. The erbium-doped fiber amplifier is a key enabling technology already emblematic of our century. Its transposition to the quantum communication world is an active subject of research showing interesting possibilities for long distance quantum cryptography \cite{QRepeater,ReviewQM}. 

The direct use of erbium-doped fiber as an optical quantum memory is extremely appealing. Nevertheless the coherence time necessary to preserve the quantumness falls in the microsecond range even at sub-Kelvin temperature \cite{Staudt2006720}. Instead of amorphous materials \cite{hastings2006controlled}, crystalline samples namely \eryso have shown remarkably long optical coherence time for solids \cite{Bottger2009}. These engaging properties are unfortunately counterbalanced by poor optical pumping dynamics. Spectral hole-burning (SHB) required by most of the quantum storage protocols is particularly challenging in erbium doped solids  \cite{Lauritzen_pumping}. This intrinsic limitation is both due to the short lifetime of the population possibly shelved in the Zeeman sublevels ($<$ 100 ms  \cite{Lauritzen_pumping}) and to the long excited state population lifetime ($\sim$ 10 ms). The ratio between the two timescales is not sufficient to obtain a good state preparation for optical thick samples. This simple experimental observation drastically bridles the implementation of quantum memories \cite{LauritzenPRA}. As an example, using the protocol named CRIB for controlled reversible inhomogeneous broadening derived from the photon-echo technique, Lauritzen obtained an efficiency of 0.25\% in \eryso \cite{LauritzenPRL} but Hedges reached 69\% in \Pras \cite{Sellars} essentially explained by different optical pumping dynamics.

We recently proposed a protocol called Revival Of Silenced Echo (ROSE). It doesn't require any state preparation \cite{ROSE}. ROSE is a sequel of the standard two-pulse echo (2PE) adapted for quantum information storage. The 2PE involves a $\pi$-pulse which has two detrimental consequences. On the one hand, it inverts the population and induces spontaneous and stimulated emission that cannot be fundamentally separated from the signal \cite{ruggiero2PE}. On the other hand, the restricting phase-matching condition imposes the beam overlap between the signal and the strong rephasing pulse rendering low noise detection difficult in practice. ROSE gets rid of both by applying two rephasing pulses bringing back the population in the ground state and relaxing the phase-matching condition.

As a descendant of the 2PE, ROSE is naturally well adapted to storage into the optically excited states whose coherence time are generally shorter than the ground state sublevels. Apparent limitation should be discussed with precaution in the case of ROSE. Without state preparation, the whole inhomogeneous broadening is available offering a large multiplexing capacity \cite{Multimode}. This latter can drastically reduce the requirements for a long memory lifetime  \cite{Collins}. In other words, for ms-storage memory, the short lifetime can be compensated by a large channel number. Such performances are accessible in \eryso with a measured 4.4 ms optical coherence time $T_2$ \cite{Bottger2009} and a typical inhomogeneous broadening of 500 MHz corresponding to $\sim$ 10$^6$ possible channels.

Even if we avoid the problem of optical pumping for the state preparation by using an appropriate protocol, \eryso is still a challenging material to work with. We did not address this issue in our previous proof-of-principle demonstration \cite{ROSE}. First of all, \yso is a birefingent biaxial crystal. The use of crossed polarized beams to isolate the signal from the rephasing pulses with a non-colinear configuration required by the ROSE may be delicate.  Second, the erbium properties are extremely anisotropic in \yso because of the low substitution site symmetry of Er$ ^{3+}$ in this compound \cite{Bottger2006bis, Bottger2008, Bottger2009}. We first present an appropriate beam, polarization and magnetic field orientation that allows the implementation of ROSE in this material. We obtain a large storage efficiency $\eta \sim$ 40\%, close to the expected maximum for the forward configuration namely 54\% \cite{ROSE} for 
\begin{equation}
\eta=(\alpha L)^{2} e^{-\alpha L}
\label{eq:eff}
\end{equation}
 
%

\eryso has been deeply studied regarding its spectroscopic and dynamic properties by B\"ottger, Sun and coworkers \cite{Bottger2006, Bottger2006bis, Bottger2008, Bottger2009}. \yso has 3 mutually perpendicular optical extinction axes called \dun, \ddeux and \baxis\!\!. Erbium substitutes two sites with transitions near 1.5 $\mu$m from the ground state $^{4}$I$_{15/2}$ to the excited state $^{4}$I$_{13/2}$. The site used for the experiment is  at 1536.48 nm, referenced as site 1 in B\"ottger's work: it has the longest coherence time. This site presents a population lifetime of around 10 ms at low temperature and an inhomogeneous linewidth of approximately 500 MHz. We use a 3 $\times$ 4   $\times$ 5 mm$^3$  crystal doped with 50 ppm grown by Scientific Materials Corp. and cooled down in a variable temperature liquid helium cryostat to 1.8 K.

We now list the different orientational constraints:


{\bf (i)} - The magnetic field $\vec{B}$ should be placed in the plane \dun-\ddeux\!\!\!. Otherwise the crystallographic site 1 splits into two magnetically inequivalent sub-classes thus reducing the optical depth \cite{Bottger2006}. More precisely in the plane \dun-\ddeux, $\vec{B}$ should ideally make an angle of 135$^\circ$ with respect to \dun\!\!. The large g factors for both sites at this angle strongly reduce the electron-spin fluctuations \cite{Bottger2009}.

{\bf (ii)} - Crossed polarizations between signal and rephasing are helpfull to isolate the echo from the strong rephasing pulses.

{\bf (iii)} - The signal and rephasing should not be colinear, neither copropagating nor counterpropagating, to further reduce the contamination of the echo by strong pulses or their specular reflections. A small angle separating quasi-counterpropating beams is desirable \cite{ROSE}. The phase-matching conditions also ensure that the 2PE potentially emitted after the first rephasing pulse is completely silenced but the ROSE echo (after the second rephasing) is emitted in the signal mode (forward).


The configuration we use (see fig. \ref{fig:Crystal}) satisfies the different conditions:

\begin{figure}[h!]
\centering
\includegraphics[height=5cm]{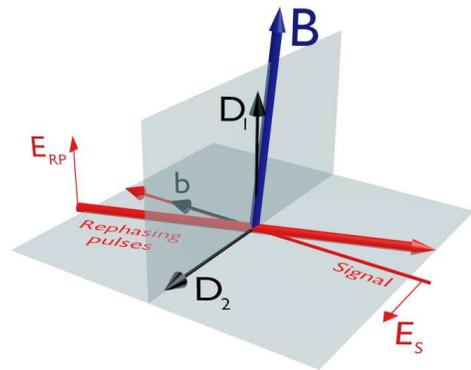}
\caption{Beams (signal and rephasing), polarizations (noted E$_\mathrm{RP}$ and E$_\mathrm{S}$ respectively) and magnetic field $\vec{B}$ orientations. \dun\!\!, \ddeux and \baxis are the \yso optical extinction axes.}
\label{fig:Crystal}
\end{figure}

We place the crystal on a rotating sample holder. We choose the  rotation axis as \!\!\baxis and the magnetic field in the cryostat is perpendicular to it by construction. We thus satisfy the first condition (i).

We take the signal beam as a reference, it propagates along \!\baxis\!\!. It polarization is made parallel to \!\ddeux in order to maximize the optical depth for the signal. The absorption is indeed anisotropic. It is weaker along \dun (by a factor of two typically). This is the reason why we polarize the rephasing pulses along \dun\!\!\!. They will be less distorted by the propagation. Cross-polarization is possible in \yso because the transition dipoles have no preferred orientation neither along \dun nor along {\bf D$_2$}. In other words, the rephasing pulses still act on the dipoles excited by the signal. We fulfill he second condition (ii).

Finally, we slightly deviate from counterpropating beam configuration for signal and rephasing by setting a few mrad angle between the two paths. The rephasing beam is then off-axis with respect to \baxis (signal) but it should be kept in the \baxis-\dun  or  \baxis-\ddeux plane otherwise it will undergo birefringence for geometrical reason. We choose the \baxis-\ddeux plane (see fig. \ref{fig:Crystal}). The last condition (iii) is satisfied.

The magnetic field amplitude should be as large as possible in order to obtain a long coherence time. In our case, the superconducting solenoid inside the cryostat is limited to 3.3 T. With the maximum field we unfortunately cannot use the configuration recommended by B\"ottger ($\vec{B}$ making a  140$^\circ$ angle with \dun) because our laser has a limited tuning range and cannot go under 1536.12 nm (see below for a complete description). For 3.3 T and $\vec{B}$ making a 135$^\circ$ angle with \dun\!\!, the transition is out of reach. Two options are possible to match the laser and the absorption line. Either we keep the 135$^\circ$ angle orientation and reduce the magnetic field or we keep the maximum field of 3.3 T and rotate the crystal to increase the angle to 160$^\circ$ typically. This second option is only possible because the g-factor is strongly anisotropic. It should induce a minor change for the  coherence time as described by B\"ottger's. We tested both options and we found that the second presents a larger $T_2$. It seems better to reduce the g-factor than the magnetic field. This statement deserves further investigations.

The complete experimental setup is described as follows. As a source, we use a commercial erbium-doped fiber laser (Koheras) with a wavelength centered at 1536.5 nm. Its temperature controlled tuning range is 0.7 nm typically. We split the output of the laser into two beams, one of them was used as the signal ($\sim$ 10 $\mu$W) and the other for the rephasing pulses. In order to get strong pulses for the rephasing beam ($\sim$~10~mW), we inject a  erbium-doped fiber amplifier. We adjust the signal and rephasing beam waists to  50 $\mu$m and 110 $\mu$m  respectively. We collect the transmitted signal corresponding to the ROSE echo using a monomode fiber and measured with an avalanche photodiode.
Both beams are shaped in time by acousto-optic modulators controlled with an Arbitrary Wave Generator Tektronix AWG520 to provide amplitude and phase control. The time sequence is essentially driven by the maximum available Rabi frequency  $\Omega_{0}$. ROSE involves adiabatic passages instead of $\pi$-pulses for the rephasing step. The pulses are called complex hyperbolic secant (CHS)\cite{De_Seze} defined as follows. They have an hyperbolic secant shape defined as a time varying Rabi frequency 
$\displaystyle
\Omega(t)=\Omega_{0}\, \mathrm{sech} \left(\beta\left(t-t_{2,3}\right)\right)
$.
The pulse is centered on $t_2$ and $t_3$  with a duration $1/\beta$. Along with the amplitude shaping, the frequency is swept around the central frequency $\omega_{0}$ with an hyperbolic tangent shape
$
\omega(t)=\omega_{0} + \mu \beta \tanh \left(\beta \left( t -t_{2,3}\right)\right)
$. 
The pulse bandwidth is then $2 \mu \beta$. They play the role of $\pi$-pulses for the standard 2PE with the advantage of higher robustness. The adiabatic condition should be maintained as $\mu \beta^2 \ll {\Omega_0}^2$ \cite{Florencia}.

In practice because the available Rabi frequency is $\Omega_0 \sim 2\pi \times 800 $ kHz, we choose $\mu$=1 and $\beta=2\pi \times 400~$kHz. It gives a storage bandwidth of 800 kHz. We set the different delays between the signal at  $t_1$ and the first rephasing pulse t$_{12}=t_2-t_1=4$ $ \mu s$, the first and the second rephasing pulse  t$_{23}=8$ $ \mu s$ to their minimum value without much overlap between the pulses. The storage time is $2 t_{23}=t_3-t_2=16$ $ \mu s$ much shorter than the expected $T_2$ (see discussion below). We repeat the experiment every 20 ms,   sufficient for the ions to go back to the ground state.

\begin{figure}[h!]
\centering
\includegraphics[width=8cm]{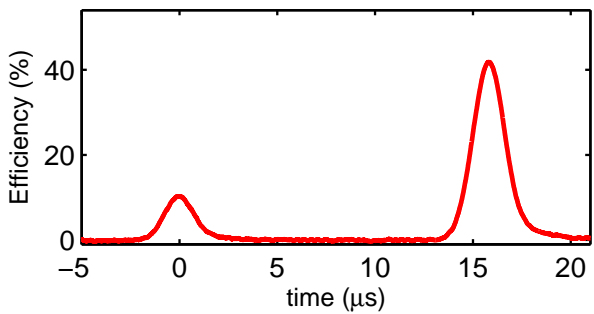}\caption{ROSE echo for $\alpha L = 2.3$. The first pulse is the transmitted signal. The echo is retrieved at  $2 t_{23}=16$ $ \mu s$ with an efficiency of 42\%.}
\label{fig:echo}
\end{figure}

With the specific orientation described previously and the laser at the minimum of its tuning range 1536.12 nm, we adjust the magnetic field by a few percent around 3.3 T to finely adjust the optical depth to $\alpha L \simeq 2$ where the efficiency should be maximum (Eq. (\ref{eq:eff})). We observe on the photodiode the partially absorbed signal at t$_{1}=0$ $ \mu s$ and the ROSE echo at  2t$_{23}=16$ $ \mu s$ (fig. \ref{fig:echo}).

To evaluate the efficiency, we independently measure the optical depth by using the SHB technique. In figure \ref{fig:echo}, the efficiency is then given by the amplitude ratio between the ROSE echo and transmitted signal multiplied by $\exp(-\alpha L)$: 42 \% for the present case.

\begin{figure}[h!]
\centering
\includegraphics[width=8.5cm]{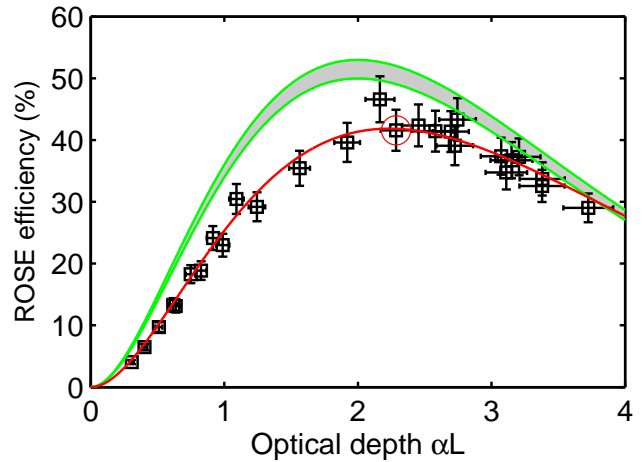}\caption{ROSE efficiency as a function of the optical depth $\alpha$L. The squares are the experimental data. The error bars essentially come from shot-to-shot fluctuations that we attribute to the laser fluctuations. The data are averaged over a few seconds. The typical deviation of  shot-to-shot fluctuations gives the error bar for both the efficiency and  $\alpha$L (from SHB). The solid lines (green and red) represents the fit to theoretically formulas (see text for details). The point corresponding to fig. \ref{fig:echo} is circled.}
\label{fig:Efficiency}
\end{figure}

The ROSE efficiency should critically depend on the optical depth $\alpha$L. To evaluate this dependency we make the crystal absorption vary by probing different frequencies within the inhomogeneous profile. Instead of detuning our Koheras laser by changing its operating temperature, we slightly change the magnetic field by 3\% around 3.3 T. It is much faster  and more reproducible than tuning the laser. A few percents change is sufficient to cover the complete 500 MHz inhomogeneous profile. Such a tiny change should not have any influence on the $T_2$. For each value of the magnetic field, we independently measure $\alpha$L by using again the SHB technique. We obtain the efficiency curve in figure \ref{fig:Efficiency}.

The experimental curve can now be compared to the theoretical prediction. To account for the decoherence, we simply modify the efficiency (Eq.  (\ref{eq:eff})) by factor  $e^{-4t_{23}/T_{2}}$. We can measure $T_{2}$ independently in different ways. First, using the ROSE time sequence, we simply change  $t_{23}$ and observe the decay of the echo. It gives $T_2=400$ $ \mu s$ . We also use a standard 2PE sequence with strong $\pi/2$ and $\pi$ pulses. We then obtain $T_2=1.4$  ms. We suspect that this discrepancy is due to instantaneous spectral diffusion, a known interaction effect between the erbium ions \cite{Bottger2006}. This certainly deserves further studies but this exceeds the scope of the present paper. Using the two $T_2$ values, we represent a confidence interval (gray area bounded by the green curves). The qualitative agreement is correct. 

We investigate the quantitative discrepancy further by assuming imperfections of the  rephasing pulses as compared to Eq. (\ref{eq:eff}), where a complete rephasing is assumed. The task is not easy because the modeling depends on the type of imperfection. We take a simpler approach by analyzing the differential equation leading to Eq.  (\ref{eq:eff}) and by introducing phenomenological coefficients. It reads as
\begin{equation}
\frac{d \mathcal{E}(z)}{dz}=- \eta_\mathrm{pop} \frac{\alpha}{2}  \mathcal{E}(z) + \eta_\mathrm{phase} e^{-\frac{2t_{23}}{T_{2}}} \alpha e^{-\frac{\alpha z}{2}}  \mathcal{S}(0) ,
\label{eq:echodiff}
\end{equation}
where  $\mathcal{S}$ and $ \mathcal{E} $  stand for the signal and the echo amplitude  respectively. We introduce $\eta_\mathrm{pop}$ and $\eta_\mathrm{phase}$ as phenomenological coefficients justified as follows. The equation is generic and describes the emission of the echo in the forward direction when the medium is in the ground state. It works both for CRIB \cite{sangouard2007} and ROSE \cite{ROSE}. It can be integrated between 0 and $L$ to obtain the efficiency as $\eta = \displaystyle\left[{ \mathcal{E}(L)}/{\mathcal{S}(0)} \right]^2$. The differential equation can be discussed in physical terms to justify the introduction of  $\eta_\mathrm{pop}$ and $\eta_\mathrm{phase}$. The first term $\displaystyle \frac{\alpha}{2}  \mathcal{E}(z)$ describes the attenuation of the echo because the medium is brought back to the ground state. To account for an imperfect return (remaining population), we introduce the factor $\eta_\mathrm{pop}$. The second term in $e^{-\frac{\alpha z}{2}} \mathcal{S}(0)$ is a source term. It represents the rephasing of the coherences previously excited by the signal. It decays as $\frac{2t_{23}}{T_2}$ during the storage time $2t_{23}$ because of decoherence. To account for an imperfect rephasing, we introduce $\eta_\mathrm{phase}$. $\eta_\mathrm{pop}$ and $\eta_\mathrm{phase}$ are clearly phenomenological because we artificially separate the return to the ground state (population) and the coherence rephasing even though they are due to the same CHS pulses. It has the advantage to provide a simple model to quantify the imperfections of the rephasing stage with ad-hoc coefficients. After integration, the efficiency can be further simplified by assuming ${\alpha L}(1-\eta_\mathrm{pop}) \ll 1$:
\begin{equation}
\eta \simeq \eta_\mathrm{phase}^2 \left(\alpha L\right)^2 e^{-\alpha L\frac{1+\eta_\mathrm{pop}}{2}}  e^{-\frac{4t_{23}}{T_{2}}}
\label{eq:effv2}
\end{equation}
For $\eta_\mathrm{pop} \rightarrow 1$ and $\eta_\mathrm{phase} \rightarrow 1$, one recovers Eq. (\ref{eq:eff}) including the decoherence term. We left these coefficients  as free parameters to fit the experimental data. We fix the value of $T_2=400$ $ \mu s$  corresponding to the observed ROSE decay. The least square optimization gives the red curve (fig. \ref{fig:Efficiency}) with $\eta_\mathrm{pop} = $ 80\% and $\eta_\mathrm{phase} = $85\%.

These values are consistent and in good agreement with the inversion quality of the CHS pulse that we measure independently to be consistently in the range 70\%-90\% \cite{ROSE}.

To conclude, we observe efficient light storage in the telecom range in \eryso. We have identified an appropriate beam configuration with respect to the crystal axis and magnetic field. On the way towards quantum storage, single photon level measurements should be done to evaluate the remaining noise by spontaneous emission. We suspect the effect of instantaneous spectral diffusion. It should be carefully investigated because it restrains the bandwidth and ultimately the multiplexing capacity of \eryso. 

The research leading to these results has received funding from the People Programme (Marie Curie Actions) of the European Union's Seventh Framework Programme FP7/2007-2013/ under REA grant agreement no. 287252 and from the national grant ANR-12-BS08-0015-02 (RAMACO).

\bibliography{ROSE_efficiency_bib}

\begin{thebibliography}{19}%
\makeatletter
\providecommand \@ifxundefined [1]{%
 \@ifx{#1\undefined}
}%
\providecommand \@ifnum [1]{%
 \ifnum #1\expandafter \@firstoftwo
 \else \expandafter \@secondoftwo
 \fi
}%
\providecommand \@ifx [1]{%
 \ifx #1\expandafter \@firstoftwo
 \else \expandafter \@secondoftwo
 \fi
}%
\providecommand \natexlab [1]{#1}%
\providecommand \enquote  [1]{``#1''}%
\providecommand \bibnamefont  [1]{#1}%
\providecommand \bibfnamefont [1]{#1}%
\providecommand \citenamefont [1]{#1}%
\providecommand \href@noop [0]{\@secondoftwo}%
\providecommand \href [0]{\begingroup \@sanitize@url \@href}%
\providecommand \@href[1]{\@@startlink{#1}\@@href}%
\providecommand \@@href[1]{\endgroup#1\@@endlink}%
\providecommand \@sanitize@url [0]{\catcode `\\12\catcode `\$12\catcode
  `\&12\catcode `\#12\catcode `\^12\catcode `\_12\catcode `\%12\relax}%
\providecommand \@@startlink[1]{}%
\providecommand \@@endlink[0]{}%
\providecommand \url  [0]{\begingroup\@sanitize@url \@url }%
\providecommand \@url [1]{\endgroup\@href {#1}{\urlprefix }}%
\providecommand \urlprefix  [0]{URL }%
\providecommand \Eprint [0]{\href }%
\providecommand \doibase [0]{http://dx.doi.org/}%
\providecommand \selectlanguage [0]{\@gobble}%
\providecommand \bibinfo  [0]{\@secondoftwo}%
\providecommand \bibfield  [0]{\@secondoftwo}%
\providecommand \translation [1]{[#1]}%
\providecommand \BibitemOpen [0]{}%
\providecommand \bibitemStop [0]{}%
\providecommand \bibitemNoStop [0]{.\EOS\space}%
\providecommand \EOS [0]{\spacefactor3000\relax}%
\providecommand \BibitemShut  [1]{\csname bibitem#1\endcsname}%
\let\auto@bib@innerbib\@empty
\bibitem [{\citenamefont {Sangouard}\ \emph {et~al.}(2011)\citenamefont
  {Sangouard}, \citenamefont {Simon}, \citenamefont {de~Riedmatten},\ and\
  \citenamefont {Gisin}}]{QRepeater}%
  \BibitemOpen
  \bibfield  {author} {\bibinfo {author} {\bibfnamefont {N.}~\bibnamefont
  {Sangouard}}, \bibinfo {author} {\bibfnamefont {C.}~\bibnamefont {Simon}},
  \bibinfo {author} {\bibfnamefont {H.}~\bibnamefont {de~Riedmatten}}, \ and\
  \bibinfo {author} {\bibfnamefont {N.}~\bibnamefont {Gisin}},\ }\href@noop {}
  {\bibfield  {journal} {\bibinfo  {journal} {Rev. Mod. Phys.}\ }\textbf
  {\bibinfo {volume} {83}},\ \bibinfo {pages} {33} (\bibinfo {year}
  {2011})}\BibitemShut {NoStop}%
\bibitem [{\citenamefont {Lvovsky}\ \emph {et~al.}(2009)\citenamefont
  {Lvovsky}, \citenamefont {Sanders},\ and\ \citenamefont {Tittel}}]{ReviewQM}%
  \BibitemOpen
  \bibfield  {author} {\bibinfo {author} {\bibfnamefont {A.~I.}\ \bibnamefont
  {Lvovsky}}, \bibinfo {author} {\bibfnamefont {B.~C.}\ \bibnamefont
  {Sanders}}, \ and\ \bibinfo {author} {\bibfnamefont {W.}~\bibnamefont
  {Tittel}},\ }\href@noop {} {\bibfield  {journal} {\bibinfo  {journal} {Nature
  Photonics}\ }\textbf {\bibinfo {volume} {3}},\ \bibinfo {pages} {706}
  (\bibinfo {year} {2009})}\BibitemShut {NoStop}%
\bibitem [{\citenamefont {Staudt}\ \emph {et~al.}(2006)\citenamefont {Staudt},
  \citenamefont {Hastings-Simon}, \citenamefont {Afzelius}, \citenamefont
  {Jaccard}, \citenamefont {Tittel},\ and\ \citenamefont
  {Gisin}}]{Staudt2006720}%
  \BibitemOpen
  \bibfield  {author} {\bibinfo {author} {\bibfnamefont {M.~U.}\ \bibnamefont
  {Staudt}}, \bibinfo {author} {\bibfnamefont {S.~R.}\ \bibnamefont
  {Hastings-Simon}}, \bibinfo {author} {\bibfnamefont {M.}~\bibnamefont
  {Afzelius}}, \bibinfo {author} {\bibfnamefont {D.}~\bibnamefont {Jaccard}},
  \bibinfo {author} {\bibfnamefont {W.}~\bibnamefont {Tittel}}, \ and\ \bibinfo
  {author} {\bibfnamefont {N.}~\bibnamefont {Gisin}},\ }\href {\doibase
  http://dx.doi.org/10.1016/j.optcom.2006.05.007} {\bibfield  {journal}
  {\bibinfo  {journal} {Optics Communications}\ }\textbf {\bibinfo {volume}
  {266}},\ \bibinfo {pages} {720 } (\bibinfo {year} {2006})}\BibitemShut
  {NoStop}%
\bibitem [{\citenamefont {Hastings-Simon}\ \emph {et~al.}(2006)\citenamefont
  {Hastings-Simon}, \citenamefont {Staudt}, \citenamefont {Afzelius},
  \citenamefont {Baldi}, \citenamefont {Jaccard}, \citenamefont {Tittel},\ and\
  \citenamefont {Gisin}}]{hastings2006controlled}%
  \BibitemOpen
  \bibfield  {author} {\bibinfo {author} {\bibfnamefont {S.~R.}\ \bibnamefont
  {Hastings-Simon}}, \bibinfo {author} {\bibfnamefont {M.~U.}\ \bibnamefont
  {Staudt}}, \bibinfo {author} {\bibfnamefont {M.}~\bibnamefont {Afzelius}},
  \bibinfo {author} {\bibfnamefont {P.}~\bibnamefont {Baldi}}, \bibinfo
  {author} {\bibfnamefont {D.}~\bibnamefont {Jaccard}}, \bibinfo {author}
  {\bibfnamefont {W.}~\bibnamefont {Tittel}}, \ and\ \bibinfo {author}
  {\bibfnamefont {N.}~\bibnamefont {Gisin}},\ }\href@noop {} {\bibfield
  {journal} {\bibinfo  {journal} {Optics communications}\ }\textbf {\bibinfo
  {volume} {266}},\ \bibinfo {pages} {716} (\bibinfo {year}
  {2006})}\BibitemShut {NoStop}%
\bibitem [{\citenamefont {B\"ottger}\ \emph {et~al.}(2009)\citenamefont
  {B\"ottger}, \citenamefont {Thiel}, \citenamefont {Cone},\ and\ \citenamefont
  {Sun}}]{Bottger2009}%
  \BibitemOpen
  \bibfield  {author} {\bibinfo {author} {\bibfnamefont {T.}~\bibnamefont
  {B\"ottger}}, \bibinfo {author} {\bibfnamefont {C.~W.}\ \bibnamefont
  {Thiel}}, \bibinfo {author} {\bibfnamefont {R.~L.}\ \bibnamefont {Cone}}, \
  and\ \bibinfo {author} {\bibfnamefont {Y.}~\bibnamefont {Sun}},\ }\href
  {\doibase 10.1103/PhysRevB.79.115104} {\bibfield  {journal} {\bibinfo
  {journal} {Phys. Rev. B}\ }\textbf {\bibinfo {volume} {79}},\ \bibinfo
  {pages} {115104} (\bibinfo {year} {2009})}\BibitemShut {NoStop}%
\bibitem [{\citenamefont {Lauritzen}\ \emph {et~al.}(2008)\citenamefont
  {Lauritzen}, \citenamefont {Hastings-Simon}, \citenamefont {de~Riedmatten},
  \citenamefont {Afzelius},\ and\ \citenamefont {Gisin}}]{Lauritzen_pumping}%
  \BibitemOpen
  \bibfield  {author} {\bibinfo {author} {\bibfnamefont {B.}~\bibnamefont
  {Lauritzen}}, \bibinfo {author} {\bibfnamefont {S.~R.}\ \bibnamefont
  {Hastings-Simon}}, \bibinfo {author} {\bibfnamefont {H.}~\bibnamefont
  {de~Riedmatten}}, \bibinfo {author} {\bibfnamefont {M.}~\bibnamefont
  {Afzelius}}, \ and\ \bibinfo {author} {\bibfnamefont {N.}~\bibnamefont
  {Gisin}},\ }\href {\doibase 10.1103/PhysRevA.78.043402} {\bibfield  {journal}
  {\bibinfo  {journal} {Phys. Rev. A}\ }\textbf {\bibinfo {volume} {78}},\
  \bibinfo {pages} {043402} (\bibinfo {year} {2008})}\BibitemShut {NoStop}%
\bibitem [{\citenamefont {Lauritzen}\ \emph {et~al.}(2011)\citenamefont
  {Lauritzen}, \citenamefont {Min\'a\ifmmode~\check{r}\else \v{r}\fi{}},
  \citenamefont {de~Riedmatten}, \citenamefont {Afzelius},\ and\ \citenamefont
  {Gisin}}]{LauritzenPRA}%
  \BibitemOpen
  \bibfield  {author} {\bibinfo {author} {\bibfnamefont {B.}~\bibnamefont
  {Lauritzen}}, \bibinfo {author} {\bibfnamefont {J.~c.~v.}\ \bibnamefont
  {Min\'a\ifmmode~\check{r}\else \v{r}\fi{}}}, \bibinfo {author} {\bibfnamefont
  {H.}~\bibnamefont {de~Riedmatten}}, \bibinfo {author} {\bibfnamefont
  {M.}~\bibnamefont {Afzelius}}, \ and\ \bibinfo {author} {\bibfnamefont
  {N.}~\bibnamefont {Gisin}},\ }\href {\doibase 10.1103/PhysRevA.83.012318}
  {\bibfield  {journal} {\bibinfo  {journal} {Phys. Rev. A}\ }\textbf {\bibinfo
  {volume} {83}},\ \bibinfo {pages} {012318} (\bibinfo {year}
  {2011})}\BibitemShut {NoStop}%
\bibitem [{\citenamefont {Lauritzen}\ \emph {et~al.}(2010)\citenamefont
  {Lauritzen}, \citenamefont {Min\'a\ifmmode~\check{r}\else \v{r}\fi{}},
  \citenamefont {de~Riedmatten}, \citenamefont {Afzelius}, \citenamefont
  {Sangouard}, \citenamefont {Simon},\ and\ \citenamefont
  {Gisin}}]{LauritzenPRL}%
  \BibitemOpen
  \bibfield  {author} {\bibinfo {author} {\bibfnamefont {B.}~\bibnamefont
  {Lauritzen}}, \bibinfo {author} {\bibfnamefont {J.~c.~v.}\ \bibnamefont
  {Min\'a\ifmmode~\check{r}\else \v{r}\fi{}}}, \bibinfo {author} {\bibfnamefont
  {H.}~\bibnamefont {de~Riedmatten}}, \bibinfo {author} {\bibfnamefont
  {M.}~\bibnamefont {Afzelius}}, \bibinfo {author} {\bibfnamefont
  {N.}~\bibnamefont {Sangouard}}, \bibinfo {author} {\bibfnamefont
  {C.}~\bibnamefont {Simon}}, \ and\ \bibinfo {author} {\bibfnamefont
  {N.}~\bibnamefont {Gisin}},\ }\href {\doibase 10.1103/PhysRevLett.104.080502}
  {\bibfield  {journal} {\bibinfo  {journal} {Phys. Rev. Lett.}\ }\textbf
  {\bibinfo {volume} {104}},\ \bibinfo {pages} {080502} (\bibinfo {year}
  {2010})}\BibitemShut {NoStop}%
\bibitem [{\citenamefont {Hedges}\ \emph {et~al.}(2010)\citenamefont {Hedges},
  \citenamefont {Longdell}, \citenamefont {Li},\ and\ \citenamefont
  {Sellars}}]{Sellars}%
  \BibitemOpen
  \bibfield  {author} {\bibinfo {author} {\bibfnamefont {M.~P.}\ \bibnamefont
  {Hedges}}, \bibinfo {author} {\bibfnamefont {J.~J.}\ \bibnamefont
  {Longdell}}, \bibinfo {author} {\bibfnamefont {Y.}~\bibnamefont {Li}}, \ and\
  \bibinfo {author} {\bibfnamefont {M.~J.}\ \bibnamefont {Sellars}},\ }\href
  {\doibase 10.1038/nature09081} {\bibfield  {journal} {\bibinfo  {journal}
  {Nature}\ }\textbf {\bibinfo {volume} {465}},\ \bibinfo {pages} {1052}
  (\bibinfo {year} {2010})}\BibitemShut {NoStop}%
\bibitem [{\citenamefont {Damon}\ \emph {et~al.}(2011)\citenamefont {Damon},
  \citenamefont {Bonarota}, \citenamefont {Louchet-Chauvet}, \citenamefont
  {Chaneli\`ere},\ and\ \citenamefont {Le~Gou\"et}}]{ROSE}%
  \BibitemOpen
  \bibfield  {author} {\bibinfo {author} {\bibfnamefont {V.}~\bibnamefont
  {Damon}}, \bibinfo {author} {\bibfnamefont {M.}~\bibnamefont {Bonarota}},
  \bibinfo {author} {\bibfnamefont {A.}~\bibnamefont {Louchet-Chauvet}},
  \bibinfo {author} {\bibfnamefont {T.}~\bibnamefont {Chaneli\`ere}}, \ and\
  \bibinfo {author} {\bibfnamefont {J.-L.}\ \bibnamefont {Le~Gou\"et}},\
  }\href@noop {} {\bibfield  {journal} {\bibinfo  {journal} {New J. Phys.}\
  }\textbf {\bibinfo {volume} {13}},\ \bibinfo {pages} {093031} (\bibinfo
  {year} {2011})}\BibitemShut {NoStop}%
\bibitem [{\citenamefont {Ruggiero}\ \emph {et~al.}(2009)\citenamefont
  {Ruggiero}, \citenamefont {Gou\"{e}t}, \citenamefont {Simon},\ and\
  \citenamefont {Chaneli\`{e}re}}]{ruggiero2PE}%
  \BibitemOpen
  \bibfield  {author} {\bibinfo {author} {\bibfnamefont {J.}~\bibnamefont
  {Ruggiero}}, \bibinfo {author} {\bibfnamefont {J.-L.~L.}\ \bibnamefont
  {Gou\"{e}t}}, \bibinfo {author} {\bibfnamefont {C.}~\bibnamefont {Simon}}, \
  and\ \bibinfo {author} {\bibfnamefont {T.}~\bibnamefont {Chaneli\`{e}re}},\
  }\href@noop {} {\bibfield  {journal} {\bibinfo  {journal} {Phys. Rev. A}\
  }\textbf {\bibinfo {volume} {79}},\ \bibinfo {eid} {053851} (\bibinfo {year}
  {2009})}\BibitemShut {NoStop}%
\bibitem [{\citenamefont {Nunn}\ \emph {et~al.}(2008)\citenamefont {Nunn},
  \citenamefont {Reim}, \citenamefont {Lee}, \citenamefont {Lorenz},
  \citenamefont {Sussman}, \citenamefont {Walmsley},\ and\ \citenamefont
  {Jaksch}}]{Multimode}%
  \BibitemOpen
  \bibfield  {author} {\bibinfo {author} {\bibfnamefont {J.}~\bibnamefont
  {Nunn}}, \bibinfo {author} {\bibfnamefont {K.}~\bibnamefont {Reim}}, \bibinfo
  {author} {\bibfnamefont {K.~C.}\ \bibnamefont {Lee}}, \bibinfo {author}
  {\bibfnamefont {V.~O.}\ \bibnamefont {Lorenz}}, \bibinfo {author}
  {\bibfnamefont {B.~J.}\ \bibnamefont {Sussman}}, \bibinfo {author}
  {\bibfnamefont {I.~A.}\ \bibnamefont {Walmsley}}, \ and\ \bibinfo {author}
  {\bibfnamefont {D.}~\bibnamefont {Jaksch}},\ }\href {\doibase
  10.1103/PhysRevLett.101.260502} {\bibfield  {journal} {\bibinfo  {journal}
  {Phys. Rev. Lett.}\ }\textbf {\bibinfo {volume} {101}},\ \bibinfo {pages}
  {260502} (\bibinfo {year} {2008})}\BibitemShut {NoStop}%
\bibitem [{\citenamefont {Collins}\ \emph {et~al.}(2007)\citenamefont
  {Collins}, \citenamefont {Jenkins}, \citenamefont {Kuzmich},\ and\
  \citenamefont {Kennedy}}]{Collins}%
  \BibitemOpen
  \bibfield  {author} {\bibinfo {author} {\bibfnamefont {O.~A.}\ \bibnamefont
  {Collins}}, \bibinfo {author} {\bibfnamefont {S.~D.}\ \bibnamefont
  {Jenkins}}, \bibinfo {author} {\bibfnamefont {A.}~\bibnamefont {Kuzmich}}, \
  and\ \bibinfo {author} {\bibfnamefont {T.~A.~B.}\ \bibnamefont {Kennedy}},\
  }\href {\doibase 10.1103/PhysRevLett.98.060502} {\bibfield  {journal}
  {\bibinfo  {journal} {Phys. Rev. Lett.}\ }\textbf {\bibinfo {volume} {98}},\
  \bibinfo {pages} {060502} (\bibinfo {year} {2007})}\BibitemShut {NoStop}%
\bibitem [{\citenamefont {B\"ottger}\ \emph
  {et~al.}(2006{\natexlab{a}})\citenamefont {B\"ottger}, \citenamefont {Sun},
  \citenamefont {Thiel},\ and\ \citenamefont {Cone}}]{Bottger2006bis}%
  \BibitemOpen
  \bibfield  {author} {\bibinfo {author} {\bibfnamefont {T.}~\bibnamefont
  {B\"ottger}}, \bibinfo {author} {\bibfnamefont {Y.}~\bibnamefont {Sun}},
  \bibinfo {author} {\bibfnamefont {C.~W.}\ \bibnamefont {Thiel}}, \ and\
  \bibinfo {author} {\bibfnamefont {R.~L.}\ \bibnamefont {Cone}},\ }\href
  {\doibase 10.1103/PhysRevB.74.075107} {\bibfield  {journal} {\bibinfo
  {journal} {Phys. Rev. B}\ }\textbf {\bibinfo {volume} {74}},\ \bibinfo
  {pages} {075107} (\bibinfo {year} {2006}{\natexlab{a}})}\BibitemShut
  {NoStop}%
\bibitem [{\citenamefont {Sun}\ \emph {et~al.}(2008)\citenamefont {Sun},
  \citenamefont {B\"ottger}, \citenamefont {Thiel},\ and\ \citenamefont
  {Cone}}]{Bottger2008}%
  \BibitemOpen
  \bibfield  {author} {\bibinfo {author} {\bibfnamefont {Y.}~\bibnamefont
  {Sun}}, \bibinfo {author} {\bibfnamefont {T.}~\bibnamefont {B\"ottger}},
  \bibinfo {author} {\bibfnamefont {C.~W.}\ \bibnamefont {Thiel}}, \ and\
  \bibinfo {author} {\bibfnamefont {R.~L.}\ \bibnamefont {Cone}},\ }\href
  {\doibase 10.1103/PhysRevB.77.085124} {\bibfield  {journal} {\bibinfo
  {journal} {Phys. Rev. B}\ }\textbf {\bibinfo {volume} {77}},\ \bibinfo
  {pages} {085124} (\bibinfo {year} {2008})}\BibitemShut {NoStop}%
\bibitem [{\citenamefont {B\"ottger}\ \emph
  {et~al.}(2006{\natexlab{b}})\citenamefont {B\"ottger}, \citenamefont {Thiel},
  \citenamefont {Sun},\ and\ \citenamefont {Cone}}]{Bottger2006}%
  \BibitemOpen
  \bibfield  {author} {\bibinfo {author} {\bibfnamefont {T.}~\bibnamefont
  {B\"ottger}}, \bibinfo {author} {\bibfnamefont {C.~W.}\ \bibnamefont
  {Thiel}}, \bibinfo {author} {\bibfnamefont {Y.}~\bibnamefont {Sun}}, \ and\
  \bibinfo {author} {\bibfnamefont {R.~L.}\ \bibnamefont {Cone}},\ }\href
  {\doibase 10.1103/PhysRevB.73.075101} {\bibfield  {journal} {\bibinfo
  {journal} {Phys. Rev. B}\ }\textbf {\bibinfo {volume} {73}},\ \bibinfo
  {pages} {075101} (\bibinfo {year} {2006}{\natexlab{b}})}\BibitemShut
  {NoStop}%
\bibitem [{\citenamefont {de~Seze}\ \emph {et~al.}(2005)\citenamefont
  {de~Seze}, \citenamefont {Dahes}, \citenamefont {Crozatier}, \citenamefont
  {Lorger{\'e}}, \citenamefont {Bretenaker},\ and\ \citenamefont
  {Le~Gou{\"e}t}}]{De_Seze}%
  \BibitemOpen
  \bibfield  {author} {\bibinfo {author} {\bibfnamefont {F.}~\bibnamefont
  {de~Seze}}, \bibinfo {author} {\bibfnamefont {F.}~\bibnamefont {Dahes}},
  \bibinfo {author} {\bibfnamefont {V.}~\bibnamefont {Crozatier}}, \bibinfo
  {author} {\bibfnamefont {I.}~\bibnamefont {Lorger{\'e}}}, \bibinfo {author}
  {\bibfnamefont {F.}~\bibnamefont {Bretenaker}}, \ and\ \bibinfo {author}
  {\bibfnamefont {J.~L.}\ \bibnamefont {Le~Gou{\"e}t}},\ }\href@noop {}
  {\bibfield  {journal} {\bibinfo  {journal} {Eur. Phys. J. D}\ }\textbf
  {\bibinfo {volume} {33}},\ \bibinfo {pages} {343} (\bibinfo {year}
  {2005})}\BibitemShut {NoStop}%
\bibitem [{\citenamefont {Pascual-Winter}\ \emph {et~al.}(2013)\citenamefont
  {Pascual-Winter}, \citenamefont {Togning}, \citenamefont {Chaneli\`ere},\
  and\ \citenamefont {Le~Gou\"et}}]{Florencia}%
  \BibitemOpen
  \bibfield  {author} {\bibinfo {author} {\bibfnamefont {M.~F.}\ \bibnamefont
  {Pascual-Winter}}, \bibinfo {author} {\bibfnamefont {R.-C.}\ \bibnamefont
  {Togning}}, \bibinfo {author} {\bibfnamefont {T.}~\bibnamefont
  {Chaneli\`ere}}, \ and\ \bibinfo {author} {\bibfnamefont {J.-L.}\
  \bibnamefont {Le~Gou\"et}},\ }\href@noop {} {\bibfield  {journal} {\bibinfo
  {journal} {New J. Phys.}\ }\textbf {\bibinfo {volume} {15}},\ \bibinfo
  {pages} {055024} (\bibinfo {year} {2013})}\BibitemShut {NoStop}%
\bibitem [{\citenamefont {Sangouard}\ \emph {et~al.}(2007)\citenamefont
  {Sangouard}, \citenamefont {Simon}, \citenamefont {Afzelius},\ and\
  \citenamefont {Gisin}}]{sangouard2007}%
  \BibitemOpen
  \bibfield  {author} {\bibinfo {author} {\bibfnamefont {N.}~\bibnamefont
  {Sangouard}}, \bibinfo {author} {\bibfnamefont {C.}~\bibnamefont {Simon}},
  \bibinfo {author} {\bibfnamefont {M.}~\bibnamefont {Afzelius}}, \ and\
  \bibinfo {author} {\bibfnamefont {N.}~\bibnamefont {Gisin}},\ }\href
  {\doibase 10.1103/PhysRevA.75.032327} {\bibfield  {journal} {\bibinfo
  {journal} {Phys. Rev. A}\ }\textbf {\bibinfo {volume} {75}},\ \bibinfo {eid}
  {032327} (\bibinfo {year} {2007})}\BibitemShut {NoStop}%
\end{thebibliography}%

%
%

%

\end{document}